\documentclass[twocolumn,letter]{jpsj3}
\usepackage{graphicx}
\usepackage{color}
\newcommand{\vect}[1]
{{\mbox{\boldmath $#1$}}}

\title{Spin-Density-Wave-Type Ordering of LaCoGe Revealed by $^{59}$Co- and $^{139}$La-Nuclear Magnetic Resonance Measurements}

\author{
Kosuke \text{Karube}$^1$\thanks{E-mail address : karube@scphys.kyoto-u.ac.jp}, 
Taisuke Hattori$^1$,  
Kenji Ishida$^{1}$\thanks{E-mail address : kishida@scphys.kyoto-u.ac.jp}, 
Nobuhiko Tamura$^2$,
Kazuhiko Deguchi$^2$, and 
Noriaki K. Sato$^2$} 

\inst{$^1$Departure of Physics, Graduate School of Science, Kyoto University, Kyoto 606-8502, Japan\\
$^2$Department of Physics, Graduate School of Science, Nagoya University, Nagoya 464-8602, Japan\\}

\abst{The low-temperature magnetic properties of LaCoGe with the tetragonal CeFeSi-type structure were investigated by $^{59}$Co- and $^{139}$La-nuclear magnetic resonance (NMR) and nuclear quadrupole resonance 
(NQR) measurements. The nuclear spin-lattice relaxation rate divided by the temperature, $1/(T_1 T)$, gradually increases with decreasing temperature and shows a kink at approximately 18 K, below which an inhomogeneous internal field appears at the Co nuclear site. 
These results indicate that antiferromagnetic ordering occurs below $T_{\rm N} \sim 18$ K.
However, an internal field was not observed at the La nuclear site below $T_{\rm N}$. Taking all NMR results into account, we conclude that spin-density-wave (SDW)-type ordering occurs, where magnetic correlations are of the checkerboard type [ $\mbox{\boldmath{$Q$}}$= ($\pi, \pi$)] in the $ab$-plane and have a long periodicity along the $c$-axis with inhomogeneous ordered moments pointing to the $c$-axis.  
}

\kword{LaCoGe, NQR, NMR}

\begin{document}
\maketitle
Cobalt intermetallic compounds exhibit interesting magnetic properties; for instance, the cubic Laves phase compound YCo$_2$ (space group: $C15$) with enhanced Pauli paramagnetism shows metamagnetic behavior with the application of a large magnetic field of $\mu_0H \sim 70$ T\cite{Ref_YCo2}, and the isostructural compound LaCo$_2$ shows weak ferromagnetism below $T_\mathrm{Curie}$ $\sim$ 130 K with a small ordered moment of 0.1 $\mu_\mathrm{B}$ \cite{Ref_LaCo2}. In addition, superconductivity was reported in the itinerant ferromagnet Y$_4$Co$_3$  \cite{Ref_Y4Co3}. Co electronic and magnetic states in these compounds have been intensively studied by $^{59}$Co-nuclear magnetic resonance (NMR) measurements.    

\begin{figure}[tbp]
\begin{center}
\includegraphics[scale=0.32]{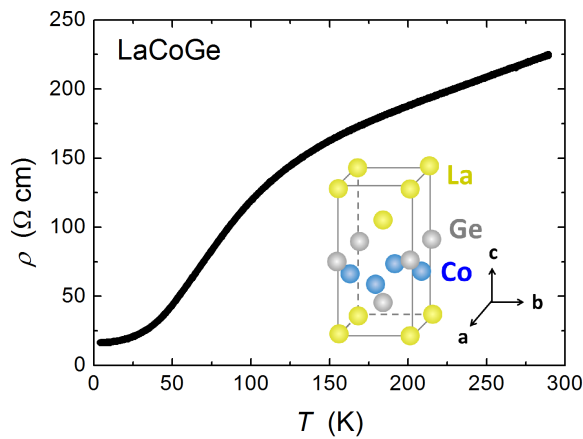}
\caption{(Color online) Temperature dependence of electrical resistivity of LaCoGe. The inset shows the tetragonal crystal structure (space group: $P4/nmm$) of LaCoGe.}
\label{LaCoGe_resistivity_CS}
\end{center}
\end{figure}

Before our study, Welter \textit{et al} and Vejpravova \textit{et al} studied LaCoGe as a reference compound of CeCoGe, since both compounds possess the same tetragonal CeFeSi-type structure (space group: $P4/nmm$) without the inversion symmetry, as shown in the inset of Fig. \ref{LaCoGe_resistivity_CS}. 
They reported that LaCoGe is a nonmagnetic compound, because no anomaly was observed down to 2 K in the magnetic susceptibility and specific heat measurements\cite{Ref_LaCoGe1,Ref_LaCoGe2}. 

However, further measurements, particularly microscopic measurements, are required in order to conclusively prove the low-temperature magnetic property of LaCoGe because it has often been observed that small amounts of impurities, which show a large Curie term in bulk susceptibility and give rise to strong scattering in transport properties, mask the intrinsic magnetic properties at low temperatures. 
We studied LaCoGe by $^{59}$Co- and $^{139}$La-NMR and nuclear quadrupole resonance (NQR) measurements, which are useful probes for obtaining intrinsic magnetic properties from a microscopic point of view.
We found a clear anomaly in the nuclear spin-lattice relaxation rate at $T_{\rm N} \sim 18$ K, below which the $^{59}$Co-NMR spectrum becomes broad owing to the appearance of a static internal field at the Co nuclear site. 
These indicate antiferromagnetic (AF) ordering occurring at $T_{\rm N}$. 
However, the internal field related to the AF ordering was not observed at La sites, which are crystallographically different from the Co sites. 
Possible magnetic correlations are discussed on the basis of the $^{59}$Co- and $^{139}$La-NMR results.   

For our NMR and NQR measurements, single-crystal samples prepared by the Czochralski method in a tetra-arc furnace were used. 
Figure \ref{LaCoGe_resistivity_CS} shows the temperature dependence of the electrical resistivity of single-crystal LaCoGe. 
The resistivity shows broad convex behavior at approximately 120 K, suggestive of the presence of strong phonon scattering and/or spin-fluctuation effects. 
However, no anomaly was observed down to 1.5 K in the resistivity measurements, which is consistent with the previous measurements\cite{Ref_LaCoGe1, Ref_LaCoGe2}.
Single-crystal samples were coarsely ground into powder ($\sim$ 100 $\mu$m) and loosely packed in a small plastic tube ($\phi \sim 5$ mm).
Two samples were prepared, i.e., a \textit{free powder} sample ($\sim 0.11$ g) and a \textit{fixed powder} sample ($\sim 0.10$ g), in which the grains are fixed with GE varnish so as not to be aligned by external fields.

\begin{figure}[tbp]
\begin{center}
\includegraphics[scale=0.23]{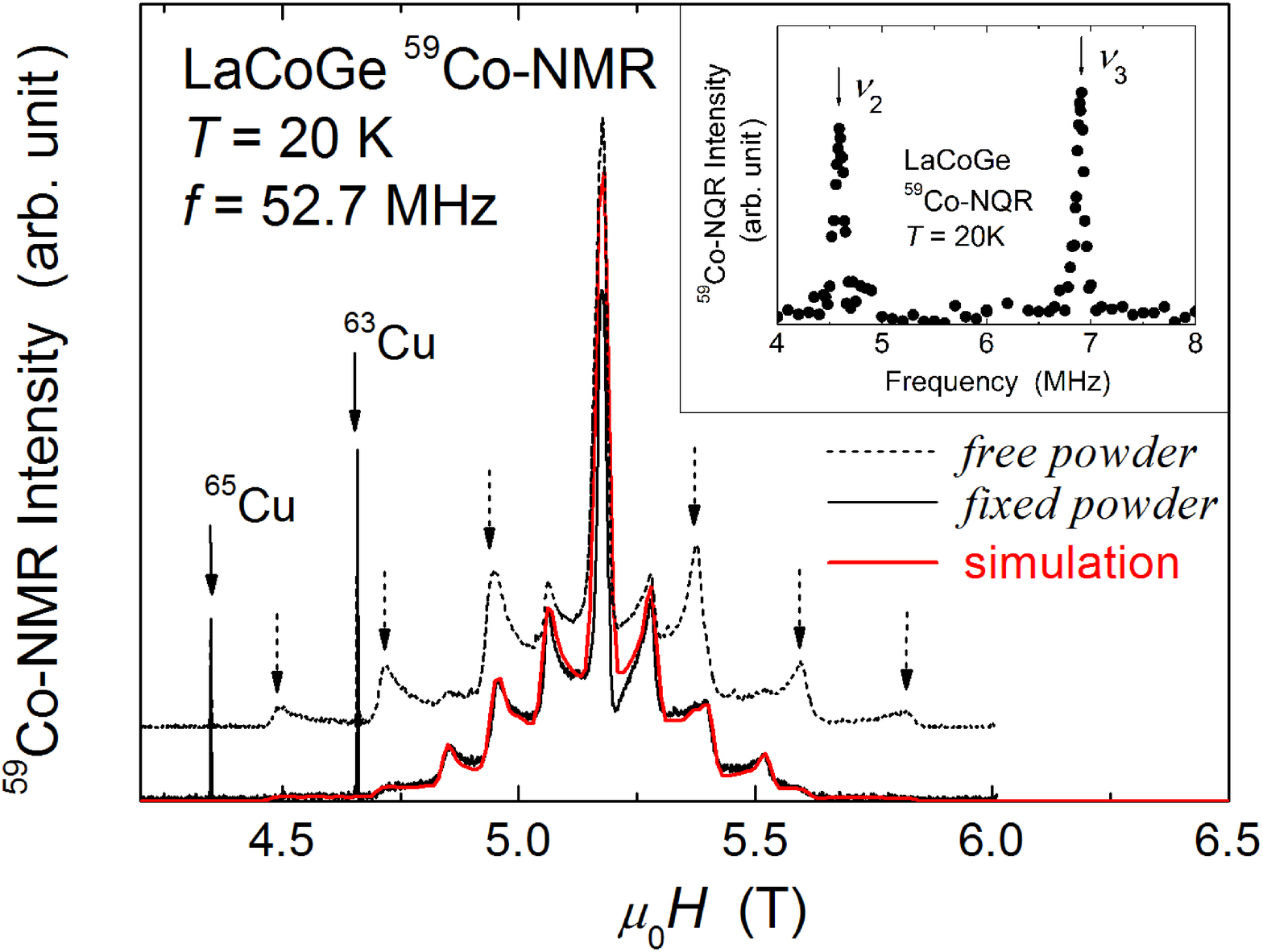}
\caption{(Color online) $^{59}$Co-NMR spectra at 20 K for \textit{free powder} (black dotted line) 
and \textit{fixed powder} (black solid line) samples. 
The red solid line is the simulation with $\nu_{zz}$ = 2.29 MHz, $\eta$ = 0, 
isotropic Knight shift $K_\mathrm{iso}$ = 1.6\%, and anisotropic Knight shift $K_\mathrm{an}$ = 0.3\%.
The peaks arising from  $^{63}$Cu and $^{65}$Cu in the NMR coil are marked by solid arrows.
Dotted arrows show the additional quadrupole satellites arising from oriented powder under an external field in 
the \textit{free powder} sample.
The inset shows the zero-magnetic-field $^{59}$Co-NQR spectrum at $T$ = 20 K. The observed peaks are $\nu_2$ = 4.59 MHz and $\nu_3$ = 6.88 MHz. 
The analysis of these peaks yields $\nu_{zz}$ = $\nu_1$ = 2.29 MHz and $\eta$ = 0.}
\label{LaCoGe_CoNMRNQRSp_T20K}
\end{center}
\end{figure}
Figure \ref{LaCoGe_CoNMRNQRSp_T20K} shows the field-swept $^{59}$Co-NMR spectra measured at 20 K
 for the \textit{free powder} and \textit{fixed powder} samples.
In general, an NMR spectrum in the presence of the electric quadrupole interaction $\mathcal{H}_{\rm Q}$ can be expressed as
\begin{eqnarray}
\mathcal{H}&=&\mathcal{H}_\mathrm{Z}+\mathcal{H}_\mathrm{Q} \nonumber\\
&=&-\gamma_\mathrm{n} \hbar (1+\mbox{\boldmath{$K$}})\mbox{\boldmath{$I$}} \cdot \mbox{\boldmath{$H$}} \nonumber\\ && + \:  \frac{\hbar\nu_{zz}}{6}\left\{ 3(I_{z}^{2}-\mbox{\boldmath{$I$}}^2)+\frac{\eta}{2}(I_{+}^{2}+I_{-}^{2}) \right\},
\end{eqnarray}
where $\mathcal{H}_\mathrm{Z}$ is the Zeeman interaction, and $\mbox{\boldmath{$K$}}$ and $\mbox{\boldmath{$H$}}$ are the Knight-shift tensor and  external-field vector, respectively. 
$\nu_{zz}$ is the frequency along the principal axis of the electric field gradient (EFG) and $\eta$ is the asymmetry parameter, defined as $\eta\equiv |V_{xx}-V_{yy}|/V_{zz}$, where $V_{ij}$ are the components of the EFG tensor.  
The typical NMR powder pattern for $I$ = 7/2 and $\eta$ = 0 was observed in the \textit{fixed powder} sample of LaCoGe.  
The $\eta = 0$ result indicates the axial symmetry of the Co sites, which is anticipated from the tetragonal crystal structure; thus, the EFG principal axis is determined to be the $c$-axis.
The obtained \textit{fixed powder} NMR spectrum can be well reproduced by the NMR powder simulation\cite{MetalicShift}, in which $\nu_{zz}$ = 2.29 MHz, $\eta$ = 0, and $K_c$ = 2.2\% and $K_a$ = 1.3\%, which are the Knight shifts along the $c$- and $a$-axes, respectively, are adopted.
In contrast, the NMR spectrum of the \textit{free powder} sample is different from that of the \textit{fixed powder} sample and 
contains additional quadrupole satellites marked by dotted arrows in Fig. \ref{LaCoGe_CoNMRNQRSp_T20K}.
Note that these additional quadrupole satellites appear at an interval two times larger than that in the \textit{fixed powder} sample.
This indicates that the magnetic easy axis in LaCoGe is parallel to the $c$-axis, which is discussed as follows.
According to the first-order perturbation calculation for the Hamiltonian $\mathcal{H}=\mathcal{H}_\mathrm{Z}+\mathcal{H}_\mathrm{Q}$ with $\eta$ = 0, quadrupole satellites appear from the central peak by the integral multiplication periodicity of the unit $\Delta H = \nu_{zz}|3\cos^2\theta -1|/(2\gamma_\mathrm{n})$, where $\theta$ is the angle between the EFG principal axis ($c$-axis in LaCoGe) and the external magnetic field. 
In the case of the \textit{fixed powder} sample, $\theta$ is randomly distributed and the number of crystals with $\theta$ = 90$^\circ$ is maximum, which yields peaks at $\Delta H_{fixed} = \nu_{zz}/(2\gamma_\mathrm{n}$). 
On the other hand, in the case of the \textit{free powder} sample, 
satellite peaks are observed at $\Delta H_{free} = 2\Delta H_{fixed}$. This is possible when the number of crystals with $\theta$ = 0$^\circ$ is maximum, indicative of the magnetic easy axis being parallel to the EFG principal axis, i.e., the $c$-axis in LaCoGe.

Zero-magnetic-field $^{59}$Co-NQR signals were observed at 4.59 and 6.88 MHz at 20 K, which are shown in the inset of Fig. \ref{LaCoGe_CoNMRNQRSp_T20K}.
These are identified as the signals arising from the $\pm 3/2 \leftrightarrow \pm 5/2$ and $\pm 5/2 \leftrightarrow \pm 7/2$ transitions, respectively, and are well understood from the above NMR parameters of $\nu_{zz}$ = 2.29 MHz and $\eta$ = 0.  
From the calculation, the signal arising from the $\pm 1/2 \leftrightarrow \pm 3/2$ transition is expected to be observed at approximately 2.29 MHz, but was not observed. This is due to the weak intensity ($I$) of the low-frequency signal ($I \propto \omega^2$) and the poor sensitivity of our NQR spectrometer in the low-frequency range.

\begin{figure}[tbp]
\begin{center}
\includegraphics[scale=0.8]{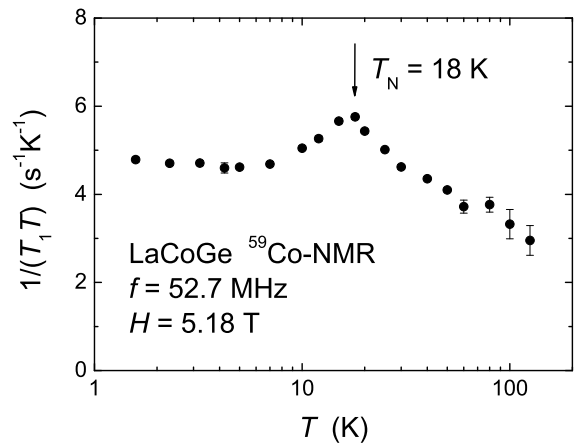}
\caption{Temperature dependence of $^{59}$Co-NMR nuclear spin-lattice relaxation rate 
divided by temperature $1/(T_1 T)$. $T_1$ was measured at the central peak and evaluated by fitting the nuclear magnetization recovery after a saturation pulse with the theoretical curve for $I$ = 7/2 and $\eta$ = 0\cite{RelaxationCurve}.
}
\label{LaCoGe_CoNMR_T1T_vsT}
\end{center}
\end{figure}
\begin{figure}[tbp]
\begin{center}
\includegraphics[scale=0.11]{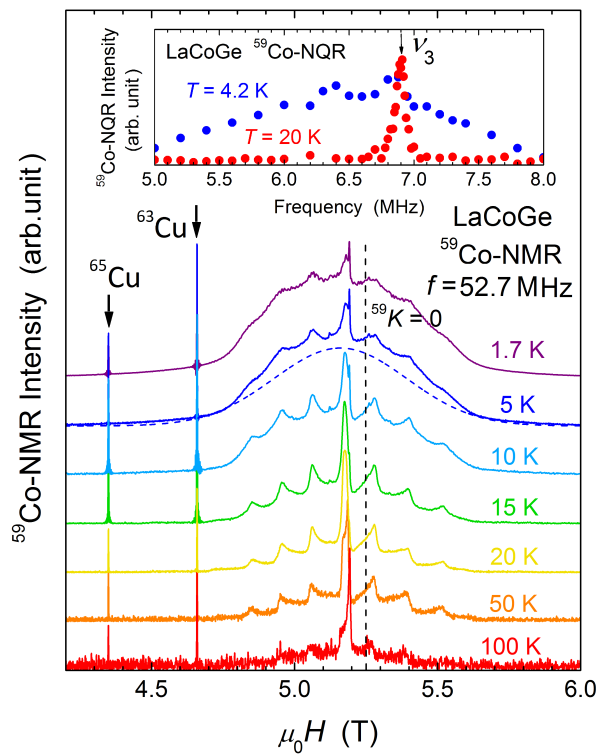}
\caption{(Color online) Temperature dependence of $^{59}$Co-NMR spectrum for the \textit{fixed powder} sample.
The intensities for each temperature are normalized by their central peaks.
The dotted blue line drawn in the 5 K spectrum is the Gaussian function $\propto \exp \left[ -\frac{\{\mu_0(H-H_0)\}^2}{2\sigma^2} \right]$ ($\mu_0 H_0$ = 0.516 T, $\sigma$ = 0.233 T), determined to be the most appropriate distribution of the internal field at 5 K.
The dotted black line $^{59}K$ = 0 shows the field where the Knight shift of $^{59}$Co is zero.
The peaks arising from $^{63}$Cu and $^{65}$Cu in the NMR coil are marked by solid arrows.
The inset shows the $\nu_3$ peak of the $^{59}$Co-NQR spectrum at $T$ = 20 and 4.2 K. 
}
\label{LaCoGe_Co-NMRNQR_SpvsT}
\end{center}
\end{figure}

The nuclear spin-lattice relaxation time $T_1$ was measured at the central peak of the $^{59}$Co-NMR spectrum of the \textit{fixed powder} sample. 
$T_1$ was evaluated by fitting the nuclear magnetization recovery after a saturation pulse with the theoretical curve for $I$ = 7/2 and $\eta$ = 0\cite{RelaxationCurve}.
The temperature dependence of $1/(T_1 T)$ is shown in Fig. \ref{LaCoGe_CoNMR_T1T_vsT}. 
$1/(T_1 T)$ gradually increases with decreasing temperature, although the Knight shift evaluated from the central peaks is nearly unchanged against temperature, as shown in Fig. \ref{LaCoGe_Co-NMRNQR_SpvsT}. 
This indicates that the AF correlation develops below 100 K, since $1/(T_1 T)$ is related to the $\mbox{\boldmath{$q$}}$-summed low-energy magnetic fluctuations [$\Sigma_ {\mbox{\boldmath{$q$}}}\chi''(\mbox{\boldmath{$q$}}, \omega \sim 0)$], and the Knight shift is proportional to the $\mbox{\boldmath{$q$}}$ = 0 static susceptibility [$\chi(\mbox{\boldmath{$q$}}=0, \omega=0)$]. 
With further decreasing temperature, $1/(T_1 T)$ shows a broad maximum at $T_\mathrm{N} \sim 18$ K, and remains constant below 7 K.
The anomaly at $T_\mathrm{N}$ and the reduction of $1/(T_1 T)$ below $T_\mathrm{N}$ are relatively small (the reduction rate is $\sim$ 20\%) in comparison with that of conventional AF compounds, indicative of a weak magnetic ordering.

The appearance of a small internal field at the Co nuclear site was detected below $T_{\rm N}$ from the $^{59}$Co-NMR spectra.
Figure \ref{LaCoGe_Co-NMRNQR_SpvsT} shows the field-swept $^{59}$Co-NMR spectra obtained at various temperatures of the \textit{fixed powder} sample. 
Below $T_\mathrm{N}$, the
original powder NMR spectrum appears to be superimposed above a broad dome-shaped background structure that gradually emerges with decreasing temperature.
However, the central peak (Knight shift) and quadrupole satellite peaks do not shift down to 1.7 K.
When the internal field with a finite value appears in the powder NMR spectrum, the NMR spectrum becomes rectangular, the width of which is double of the internal field\cite{AF-NMRSp1,AF-NMRSp2}.      
The NMR spectra below $T_{\rm N}$ indicate that the internal field is distributed from zero to a finite value, and that the average of the distributed internal field is $\sim 0.23$ T, determined from the standard deviation ($\sigma$) of the Gaussian function fitting the $^{59}$Co-NMR spectrum at 5 K, as shown in Fig. \ref{LaCoGe_Co-NMRNQR_SpvsT}.
The distribution of the internal field is also consistent with the broadening of the zero-magnetic-field $^{59}$Co-NQR spectrum shown in the inset of Fig. \ref{LaCoGe_Co-NMRNQR_SpvsT}. Thus, the AF ordering is not a field-induced one, but is present even at zero magnetic field.
\begin{figure}[tbp]
\begin{center}
\includegraphics[scale=0.8]{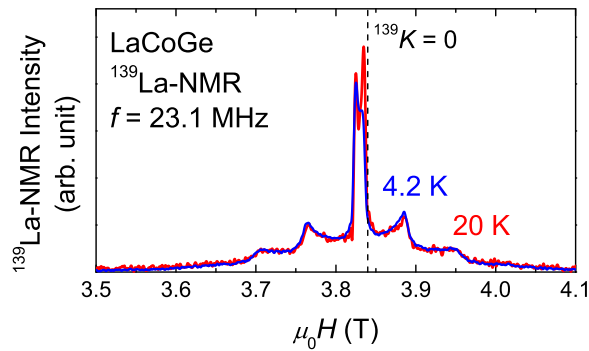}
\caption{(Color online) $^{139}$La-NMR spectra at 20 and 4.2 K for the \textit{fixed powder} sample.
The intensities for each temperature are normalized with their central peak.
The dotted line $^{139}K$ = 0 shows the field where the Knight shift of $^{139}$La is zero.}
\label{LaCoGe_LaNMR_spectrum_vsT}
\end{center}
\end{figure}

In order to obtain information about the magnetic structure below $T_{\rm N}$, we performed $^{139}$La-NMR measurement, since the La atom is located at the center of the square formed by four Co atoms.  
Figure \ref{LaCoGe_LaNMR_spectrum_vsT} shows the field-swept $^{139}$La-NMR spectra obtained at 20 and 4.2 K for the \textit{fixed powder} sample.
Unlike the above $^{59}$Co-NMR spectrum, the $^{139}$La-NMR powder spectrum is unchanged below $T_{\rm N}$, indicative of the cancellation of the internal field arising from the Co sites. 
This gives a strong constraint on the magnetic structure, as discussed below.

\begin{figure}[tbp]
\begin{center}
\includegraphics[scale=0.5]{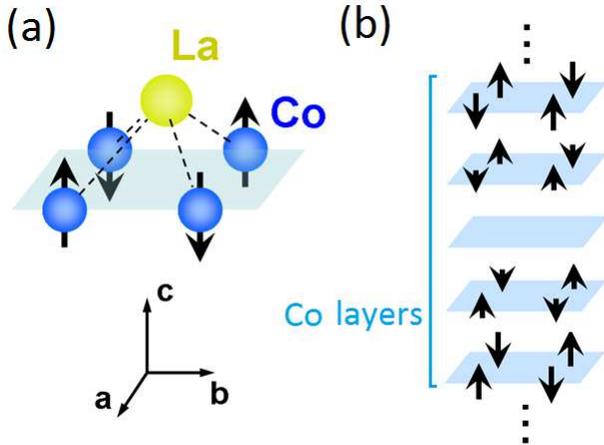}
\caption{(Color online) Schematic figures of the possible AF alignment of Co-3$d$ spins.
(a) One La atom and four nearest-neighbor Co atoms form a square pyramid structure.
In order that the internal fields arising from the Co sites are canceled out at the La site, 
the Co-3$d$ moment should be directed along the $c$-axis with the checkerboard-type AF correlation in the $ab$-plane 
and (b) the size of the moments should be distributed from zero to a finite value along the $c$-axis.
}
\label{LaCoGe_MS}
\end{center}
\end{figure}
As shown in Fig. \ref{LaCoGe_MS}(a), the La site has a symmetrical position, and the La-Co bonds form a square pyramid.
Here, we discuss the possible AF alignment of Co-3$d$ spins in this square pyramid structure.
Following the discussion of the analyses of the $^{75}$As-NMR spectra in the AF states of BaFe$_2$As$_2$\cite{Tensor1} and LaFeAsO$_{1-x}$F$_{x}$\cite{Tensor2}, 
we discuss the possible magnetic structure of LaCoGe. When the magnitudes of the four Co-3$d$ AF moments are the same, 
the internal field at the La site can be described as follows;
\begin{eqnarray}
\vect{H}^\mathrm{La}_\mathrm{int}=\displaystyle\sum_{i=1}^{4}\vect{\mathrm{A}}_{i}\cdot\vect{m}^\mathrm{Co}_{i}
=\vect{\mathrm{A}}\cdot\vect{m}^\mathrm{Co},
\end{eqnarray}
where $\vect{m}^\mathrm{Co}_{i}$ is the $i$th Co-3$d$ moment and $\vect{\mathrm{A}}_{i}$ is the hyperfine coupling tensor between the La nucleus and the $i$th Co-3$d$ moment. $\vect{m}^\mathrm{Co}$ is one of the four $\vect{m}^\mathrm{Co}_{i}$, and $\vect{\mathrm{A}}$ is the hyperfine coupling tensor contributing to the sum of the four $\vect{m}^\mathrm{Co}_{i}$ and is expressed by the orthorhombic notation as follows;
\begin{eqnarray}
\vect{\mathrm{A}}=\left(
\begin{array}{ccc}
0 & C & S_a \\
C & 0 & S_b \\
S_a & S_b & 0 
\end{array}
\right).
\end{eqnarray}
Here, $S_{a[b]}$ is a nonzero element when the stripe-type AF alignment along the $a [b]$-axis, $\vect{Q}$ = $(\pi, 0) [(0, \pi)]$, is realized, and $C$ is a nonzero element when the checkerboard-type AF alignment in the $ab$-plane, $\vect{Q}$ = $(\pi,\pi)$, is realized.
Other AF alignments, such as those of the spiral type (moments rotating by 90$^\circ$ in the $ab$-plane), are excluded because they provide finite internal fields at the La site. 
To cancel the internal field at the La site ($\vect{H}^\mathrm{La}_\mathrm{int}$ = 0), the following two cases of magnetic structure are possible. 
\begin{description}
\item{(i)} Stripe-type AF alignment along the $a [b]$-axis ($S_a$ $\neq$ 0, $S_b$ = $C$ = 0)[($S_b$ $\neq$ 0, $S_a$ = $C$ = 0)] with the moments pointing to the $b [a]$-axis ($\vect{m}^\mathrm{Co}$ = $(0, m^\mathrm{Co}_b, 0)$[$(m^\mathrm{Co}_a, 0, 0)$]).
\item{(ii)} Checkerboard-type AF alignment in the $ab$-plane ($S_a$ = $S_b$ = 0, $C$ $\neq$ 0) with the moments pointing to the $c$-axis ($\vect{m}^\mathrm{Co}$ = $(0, 0, m^\mathrm{Co}_c)$).
\end{description}
We conclude that the most promising magnetic structure of LaCoGe is (ii), as shown in Fig. \ref{LaCoGe_MS}(a), from the experimental finding that the magnetic easy axis is the $c$-axis.
Therefore, the distribution of the ordered moment, which is revealed by the $^{59}$Co-NMR/NQR spectra, should be along the $c$-axis, and the $\mbox{\boldmath{$Q$}}$-vector along the $c$-axis should have a long periodicity as shown in Fig. \ref{LaCoGe_MS}(b); thus, the magnetic state below $T_\mathrm{N}$ is interpreted to be of the spin-density-wave (SDW) type.   
We emphasize that LaCoGe is a rare Co-3$d$ compound exhibiting SDW-type ordering, since most Co-3$d$ compounds exhibit ferromagnetic ordering\cite{Ref_YCo2,Ref_LaCo2,Ref_Y4Co3}.

In summary, from the $^{59}$Co-NQR/NMR measurements of LaCoGe, we found that the AF ordering occurs at $T_{\rm N} \sim 18$ K, below which inhomogeneous internal fields appear at the Co site.
This AF ordering is considered to be weak, since the anomaly of $1/(T_1 T)$ at $T_{\rm N}$  and the internal field appearing below $T_{\rm N}$ are small.  
However, an internal field arising from the Co ordered moments was not observed at the La nuclear site below $T_{\rm N}$. 
Taking all NMR results into account, we conclude that SDW-type ordering occurs, where magnetic correlations are of the checkerboard type [ $\mbox{\boldmath{$Q$}}$= ($\pi, \pi$)] in the $ab$-plane and have a long periodicity along the $c$-axis with inhomogeneous ordered moments pointing to the $c$-axis.
Neutron diffraction and scattering experiments on LaCoGe are desired to determine the $\mbox{\boldmath{$Q$}}$-vector directly in the AF-ordered and  paramagnetic states, and to estimate the magnitude of the ordered moments.

We are grateful to S.~Yonezawa and Y. Maeno for fruitful discussions. 
This work is supported by a Grant-in-Aid for Scientific Research on Innovative Areas ``Heavy Electrons" (No.~20102006) from MEXT, for the GCOE Program ``The Next Generation of Physics, Spun from Universality and Emergence" from MEXT, and for Scientific Research from JSPS. 
One of the authors (KK) was financially supported by a JSPS Research Fellowship.

\end{document}